\documentclass{aa}

\usepackage{graphicx, epstopdf, multicol, blindtext, isotope,mathptmx}
\usepackage{txfonts, color, caption, subcaption, paralist, longtable}
\usepackage[normalem]{ulem}
\usepackage{natbib}
\bibpunct{(}{)}{;}{a}{}{,}

\newcommand{\Msun}{\,\mathrm{M}_\odot}

\newcommand{\kms}{\,$\mathrm{km}\,\mathrm{s}^{-1}$}

\begin{document}

   \title{Predicted spatial and velocity distributions of ejected companion stars of helium accretion-induced thermonuclear supernovae}

   \author{P. Neunteufel
   	\inst{1}
   	\and
   	M. Kruckow
   	\inst{2}
   	\and
   	S. Geier
   	\inst{3}
   	\and
   	A.S. Hamers
   	\inst{1}
   }

   \institute{\inst{1}Max Planck Institut f\"ur Astrophysik,
              Karl-Schwarzschild-Straße 1, 85748 Garching bei M\"unchen, Germany\\
              \inst{2}Yunnan Observatories, Chinese Academy of Sciences, 
              Kunming 650011, China\\
              \inst{3}Institut f\"{u}r Physik und Astronomie, Universit\"{a}t Potsdam, 
              Haus 28, Karl-Liebknecht-Str. 24/25, D-14476 Potsdam-Golm, Germany\\
   \email{pneun@mpa-garching.mpg.de} }

   \date{Received (month) (day), (year); accepted (month) (day), (year)}
\abstract
   {Thermonuclear supernovae (SNe), a subset of which are the highly important SNe Type\,Ia, remain one of the more poorly understood phenomena known to modern astrophysics. In recent years, the single degenerate helium (He) donor channel, where a white dwarf star (WD) accretes He-rich matter from a hydrogen-depleted companion, has emerged as a promising candidate progenitor scenario for these events. An unresolved question in this scenario is the fate of the companion star, which would be evident as a runaway hot subdwarf O/B stars (He sdO/B) in the aftermath of the SN event.}
   {Previous studies have shown that the kinematic properties of an ejected companion provide an opportunity to closer examination of the properties of an SN progenitor system. However, with the number of observed objects not matching predictions by theory, the viability of this mechanism is called into question. In this study, we first synthesize a population of companion stars ejected by the aforementioned mechanism, taking into account predicted ejection velocities, inferred population density in the Galactic (Gal.) mass distribution and subsequent kinematics in the Gal. potential. We then discuss the astrometric properties of this population.}
   {We present $10^{6}$ individual ejection trajectories, numerically computed with a newly developed, lightweight simulation framework. Initial conditions are randomly generated, but weighted according to Gal. mass density and ejection velocity data. We then discuss the bulk properties (Gal. distribution and observational parameters) of our sample.}
   {Our synthetic population reflects the Gal. mass distribution. A peak in the density distribution for close objects is expected in the direction of the Gal. centre. Higher mass runaways should outnumber lower mass ones. If the entire considered mass range is realized, the radial velocity distribution should show a peak at 500\kms. If only close US\,708 analogues are considered, there should be a peak at ($\sim750-850$)\kms. In either case, US\,708 should be a member of the high-velocity tail of the distribution.}
  % conclusions heading
  {We show that the puzzling lack of confirmed surviving companion stars of thermonuclear SNe, though possibly an observation-related selection effect, may indicate a selection against high mass donors in the SD He donor channel.}
   
   \titlerunning{Companion star kinematic properties}
\maketitle

\section{Introduction} \label{sec:introduction}
The study and theoretical characterization of thermonuclear supernovae (SNe) is a long-standing issue in modern astrophysics. Chief among these are SNe of Type\,Ia, which, distinguished by their relatively uniform light curves and well defined relation between peak luminosity and light curve shape, have been used successfully to measure the rate and acceleration of cosmic expansion \citep[Nobel Prize 2011, see][]{PAG1999}. Observationally, this class of stellar explosions is characterized by lack of evidence for the presence of hydrogen and is associated with a number of related events. These related events include SNe of Type\,Iax \citep{LFC2003,FCC2013} and, possibly, certain calcium-enhanced Type\,Ib \citep{WSL2011,NYL2017}. Thermonuclear SNe derive their name from their assumed origin as thermonuclear explosions of white dwarf stars (WDs), the degenerate remnants of low to low-intermediate mass ($\lesssim 8\Msun$) main sequence (MS) stars. Here, detonation is triggered by mass transfer from a companion star of some description. This explosion mechanism is distinguished, on the theoretical side, from core collapse SNe, generally triggered by the gravitational collapse of the core of higher-mass ($8\Msun \lesssim M \lesssim 40\Msun$) MS stars \citep[see][for a review]{SEWB2016} and, at the highest masses, pair instability SNe (PISN) become a possibility \citep[see][for reviews]{HW2002}. Note that PISN are also thermonuclear in nature, but without involvement of a WD.

Thermonuclear SNe (of the type under consideration here) are generally categorized, depending on the structure of their progenitor systems, into two distinct channels: Double degenerate (DD), where the companion is another WD, and single degenerate (SD), where the companion is a non-degenerate star, such as a MS star \citep[see][for reviews]{HN2000,R2020}.

An open question in studies of the SD channel is that of the nature of the non-degenerate companion, with stars on the MS, red giant branch (RGB), hydrogen depleted stars (hereafter: He-stars) either during or after their core helium (He) burning phase having been proposed and studied by different authors \citep[e.g.][]{N1982a,WK2011,BSB2017}.

However, regardless its structure, the donor star is expected to survive the accretor's detonation and, hence, the SN, in all viable SD scenarios \citep[see, e.g.][]{LMSW2015,LZ2020}. 
If, in the aftermath of a thermonuclear SN, the accretor is completely disrupted, then the former donor star is, according to conservation of orbital angular momentum, flung away from the location of the progenitor system at a velocity, due to ejecta interaction, slightly higher than its pre-detonation orbital velocity, as dictated by the involved masses and the binary orbital separation \citep{BWB2019}. Discarding certain super-Chandrasekhar-mass mechanisms \citep[see, e.g.][]{DVC2011}, which lie outside the parameter space considered in this study, detonation is presumably triggered by mass transfer from the non-degenerate donor star, the donor star can be assumed to be filling its Roche lobe at the point of detonation. The final attainable orbital separation, and thus runaway velocity, is thus dictated by the radius of the donor star. Consequently, the lowest runaway velocities are expected in progenitor binaries containing a low mass RGB donor, followed by He-giants and typical MS stars. The highest runaway velocities in the SD channel are expected to require a low-mass, hydrogen depleted star \citep{JWP2009,N2020}. If the plethora of channels leading to thermonuclear SNe, independent of their observational characteristics as canonical SNe Ia or related type, include the single degenerate He-donor channel in any significant capacity, there should exist a population of surviving donor stars that US\,708 is a member of. In this study, we predict the spatial and velocity distribution of this hypothetical population and attempt to both explain and draw conclusions from the current absence of US\,708 analogues from the observational record.

For most reasonable terminal mass combinations, the ejected donor star in this scenario is expected to move at sufficiently high velocities post-ejection to become unbound from the Galaxy. Such unbound stars are, in addition to their other attributes, commonly called hypervelocity stars (HVS). While many observed HVS, particularly the more massive specimens \citep[][the latter being a recent review]{H1988,B2015} are thought to originate from encounters of binary star systems with the supermassive black hole at the centre of the Galaxy, with one object \citep{KBD2020} having been traced back to this origin with high probability. HVS originating from the Galactic (Gal.) centre have also attracted attention as tracers of the Gal. mass distribution \citep{YM2007}. A substantial body of work \citep[e.g.][]{BGK2009,LZY2010,RKS2014} has considered the kinematic properties of these objects. Possible origins as former companion stars have also been considered \citep{ERR2020} and the search for HVS of all masses is ongoing \citep{RIH2020,LLL2020}.
However, at the low-mass end of the mass distribution, a number of objects have been detected \citep{GFZ2015,VNK2017,SBG2018}, likely ejected in thermonuclear SN events. Of these, the hot subdwarf US\,708 is of particular interest, as its current structure, composition and velocity suggest it as a likely product of the He donor SD channel \citep[][hereafter G15 and N20 respectively]{GFZ2015,N2020}.

Unlike surviving companions resulting from different candidate progenitor channels, US\,708 analogues (i.e. hypervelocity hot subdwarfs) are, again due to their composition and velocity, relatively clearly identifiable, if not easily detectable.
However, with direct observation of clear thermonuclear SN progenitor systems not forthcoming, analysis of their surviving ejected companions would offer an avenue of investigation into the terminal state of thermonuclear SN progenitor systems not available otherwise.
An open question in this regard is whether the serendipitous detection of US\,708, which can be described as being sui generis, indeed suggests a, as of yet undetected, population US\,708 analogues.
Belonging to the class of He-enriched hot subdwarf O (sdO) stars, US\,708 is currently the only known unbound hypervelocity He-star. To date about 6000 hot subdwarfs are known \citep{G2020,H2016}, which are found in all Galactic populations \citep{LNPL2020}. Several fast, yet still bound hot subdwarfs  have been discovered \citep{THG2011,NZI2016,ZHG2017} and the halo population contains many more sdO/Bs with quite extreme, i.e. marginally bound, orbits \citep{LNPL2020}. It is therefore not straightforward to distinguish between bound runaways and halo stars based on their kinematics. 

This paper is aimed at predicting the most likely kinematic properties of such a hypothetical population of US\,708 analogues, taking into account, as much as possible, uncertainties in the areas of explosion mechanism, terminal component mass and predicted scatter in runaway velocities. Movement in the Gal. potential and the Gal. density distribution are fully considered.

\section{Physical assumptions} \label{sec:assumptions}
As stated above, the SN channel under consideration here is the SD He-donor channel. More specifically, we consider the case of the mass donor being a low-mass hydrogen depleted, core-He burning star (i.e. a He sdO/B). This channel was studied in detail by N20, who showed that the ejection velocities of runaways produced with this mechanism are largely dictated by the terminal donor and accretor masses, not the history of the binary following the emergence of its terminal structure as a system consisting of a He subdwarf and a WD, nor the amount of transferred material.
The ejection locations in this study are assumed to lie in the Gal. disk and outside of clusters and close associations, but we do not expect our results to change significantly if this assumption is relaxed (See detailed discussion in App.\,\ref{app:dyn-friction}).

This study relies on a combination of the results of detailed stellar evolution calculations with stellar kinematics (as detailed below), which does allow a large number of experiments to be performed while being computationally relatively inexpensive. 
\subsection{Initial conditions of the run}
The initial conditions for our population consist of the ejection velocity for each ejected runaway subdwarf, which is a function of its mass, the ejection location, i.e. the location of the progenitor binary within the Gal. mass distribution, and the ejection direction, i.e. the direction, with respect to Gal. rest frame coordinates, in which the runaway is ejected. Each of these parameters is generated randomly (see App.\,\ref{app:initial conditions}). We briefly note that velocities, as a function of mass, are chosen based on the velocity spectra presented by N20 (see App.\,\ref{app:velocity spectra} for details), all ejections are assumed to originate in the Gal. plane (i.e $z_\mathrm{Gal}=0$) and the initial radial distance from the Gal. centre is weighted (see App.\,\ref{app:Galactic-density} for details) according to the density distribution obtained by \citet{IWT2013}.
\subsection{Kinematic simulation} 
Based on our randomly generated set of initial conditions, we use the SHyRT framework (See App.\,\ref{app:SHyRT}) to calculate the resulting runaway trajectories. Specifically, using the given model Gal. potential, we calculate the average orbital velocity at the point of ejection and, taking into account the ejection direction and ejection velocity with respect to the centre of mass of the progenitor binary, we calculate the Gal. rest frame velocity of the ejected star. We then allow each ejected star to move in the Gal. potential (see App.\,\ref{app:Galactic-potential} for details) for a span of $300$\,Myr, which is somewhat higher than the maximum nuclear timescale of a low mass ($\sim 0.45\Msun$) He-star (see App.\,\ref{app:lifetimes}). 
\subsection{Synthesis of the runaway population} 
The resulting trajectory data is then post-processed to generate the expected runaway population. This is accomplished by assigning each trajectory a timestamp corresponding to time intervals mimicking the inferred Gal. SN Ia rate. The ejection interval is assumed to be constant for the duration covered by the simulation at  $3\cdot10^{-3}\,\mathrm{yr}^{-1}$ \citep{CTT1997}. This assumption is at odds with SN Ia rates following a delay time distribution conforming to $\approx1/t$ \citep{RBF2009,MMF2011}. Imposing this condition in a trial run leads to a decrease of the ejection rate over the time frame of our calculations on the order of $\sim 2-3 \%$ and has a negligible effect on our results. A similar argument applies to the question of the Galactic star formation history. Thermonuclear SNe resulting from the channel under discussion here are generally thought to occur less then one\,Gyr \citep[see e.g.][who argue for $<800\,\mathrm{Myr}$]{WJH2013} after starburst. In the relevant time frame of about one\,Gyr, the Galactic star formation rate has been approximately constant \citep{FRK2004,XSL2018}. As our trajectory calculations use a basic time step of one Myr, intermediate positions are interpolated from the data. Furthermore, trajectories are truncated according the nuclear timescales of typical He-stars of the respective masses (see App.\,\ref{app:lifetimes}). Time-dependent evolution of the population can be expected. Truncation of the He-star lifetime leads to an equilibrium between objects being removed from the sample and replenishment of the population through continuous ejection. Close US\,708 analogues (with masses in the range $0.35\Msun \leq M \leq 0.4\Msun$), due to their lifetimes exceeding 300\,Myr, are not affected by this prescription. These objects will be ejected with velocities exceeding 800\kms and can be assumed to have moved sufficiently during that time to likely be inaccessible to observation. 

\section{Results} \label{sec:results}

The reader should note that we base our analysis on two different premises: First, we assume that the population consists of runaways, in a flat distribution, of the entire range of masses ($0.2\Msun \leq M \leq 0.8\Msun$) and, second, a truncated distribution with only runaways of masses $0.35\Msun \leq M \leq 0.4\Msun$ taken into consideration. The latter is chosen to represent close analogues of the observed star US\,708, which was previously inferred to have a current mass in the range $0.3\Msun \leq M \leq 0.4\Msun$ (G15, N20) with a terminal accretor mass corresponding to the assumed $1.4\Msun$ in this study. This, referring to \citet{BWB2019}, leads to a pre-detonation mass-estimate of $\sim0.35\Msun \leq M \leq 0.4\Msun$. The former distribution will be referred to as the "full sample", while the latter will be referred to as the "reduced sample". The total number of objects in the reduced sample (encompassing two out of a total of 13 discrete terminal mass values, compare App.\,\ref{app:velocity spectra}) is about 15\% that of the full sample.
Note also that the main goal here is not to provide predictions of the total number of observable objects, as uncertainties in the production rate of the progenitor systems, as well as detonation mechanisms, inhibiting unambiguous identification of spectrally classified SNe and progenitor systems, are still too great to overcome \citep[see, however,][who estimate the rate of He donor induced SNe to account for about a third of the inferred Gal. SN\,Ia rate]{WJH2013}. Instead we present relative normalized number densities, corresponding to a probability distribution for the most likely observable characteristics.

\subsection{The galactocentric distribution} \label{sec:galactocentric}
\begin{figure*}
	\centering 
	\input{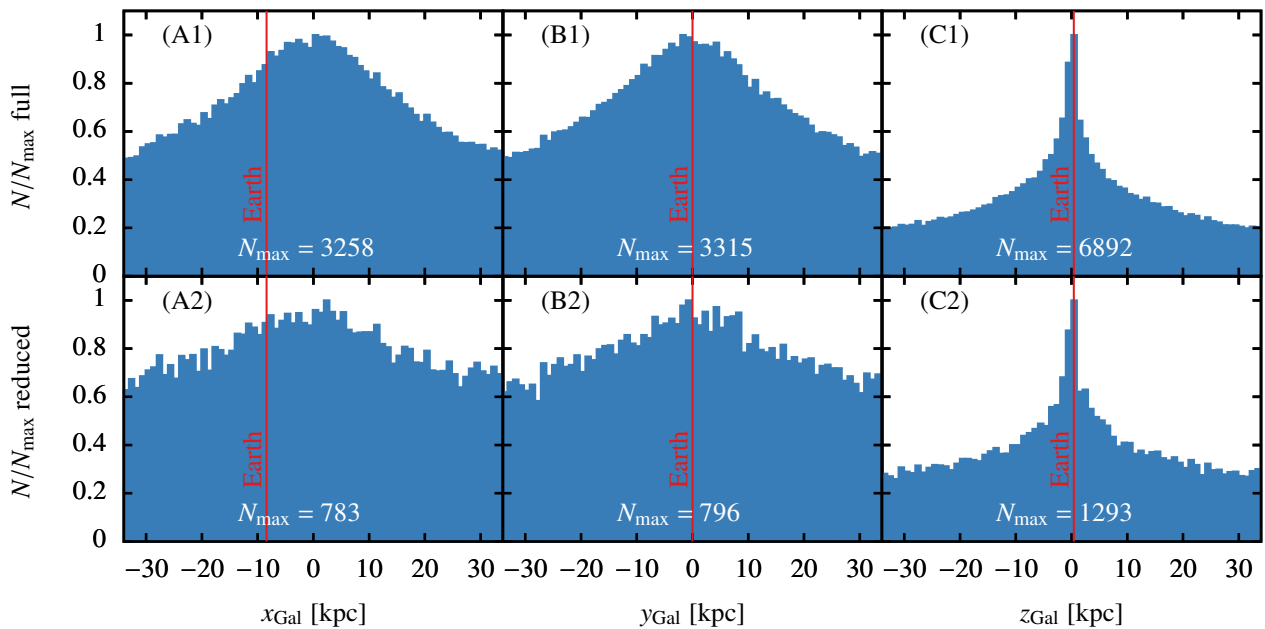}
	\vspace{0.3cm}
	\caption{Normalized number density distributions of the synthetic population in Gal. Cartesian coordinates as labelled. Upper panels show the entire mass range (full sample) of runaway masses, lower panels show the same for the US\,708-analogue subsample (reduced sample). The relative position of Earth along each axis is as indicated. Note that the distribution extends beyond the plotted range in all subplots, the depicted range chosen for clarity.} \label{fig:gal-coord}
\end{figure*}

We present results for the galactocentric distribution of our synthetic population in Fig.\,\ref{fig:gal-coord}. As shown in Figs.\,\ref{fig:gal-coord}A and B, the resulting distribution is essentially cylindrically symmetric. The apparent peak in the density distribution around the Gal. centre ($x_\mathrm{Gal} = 0$, $y_\mathrm{Gal} = 0$) is a result of the, somewhat dispersed, initial distribution conforming to the Gal. density distribution. A noticeable peak in the density distribution at the location of the Gal. disc ($z_\mathrm{Gal} = 0$) is apparent, reflecting both the initial distribution (projected onto the z-axis) and deflection of the random ejection direction due to Gal. rotation.
We conclude at this point that, neglecting instrument related selection effects, observations focused toward the Gal. centre will yield a higher probability of new discoveries. Further, observations should (assuming that the entire mass range is encountered in reality), yield a higher ratio of more massive runaways close to the Gal. centre and the Gal. disc, numbers of less massive (and therefore longer lived), runaways being reduced as these objects are ejected. Concurrently, more massive runaways will encounter the end of their He burning lifetime at shorter distances from their ejection locations both due to their shorter lifetimes and their slower ejection velocity.

\subsection{The observational distribution} \label{sec:observational}
\begin{figure*}
	\centering 
	\input{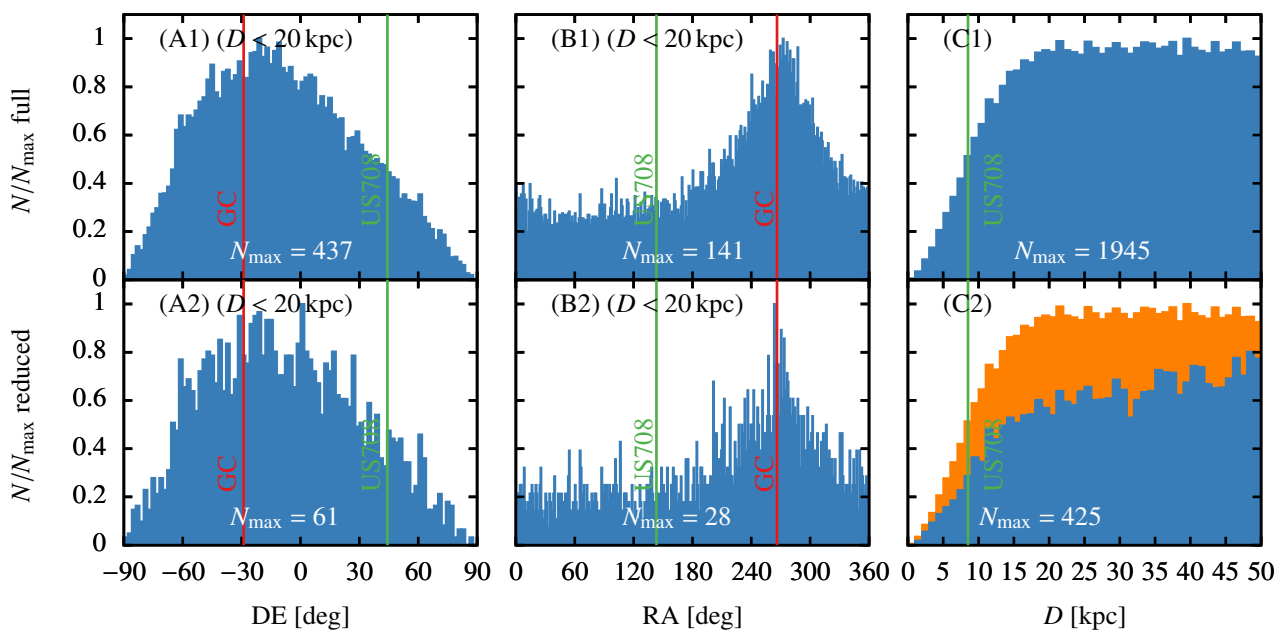}
	\vspace{0.3cm}
	\caption{Observational characteristics of our synthetic population in right ascension (RA), declination (DE) and heliocentric distance ($D$). Upper panels show the entire mass range (full sample) of runaway masses, lower panels show the same for the US\,708-analogue subsample (reduced sample). The full distribution $D$-distribution (panel C1) is shown in orange in panel C2 for comparison.  Relative positions of Gal. centre (GC) and US\,708 are given for reference. Note that, in RA and DE, the sample is restricted to objects closer than 20\,kpc. See App.\,\ref{app:full-D} for an extended plot of the distance distribution (C1 and C2).} \label{fig:obs-coord-RADE}
\end{figure*}

The celestial coordinate distribution in right ascention (RA) and declination (DE) of our sample, as well as the distance, are summarized in Fig.\,\ref{fig:obs-coord-RADE}. Note that here we focus on "closer" objects (Distance $D<20\mathrm{kpc}$) in both sub-figures (A) and (B). We note a significant peak both in RA and DE (RA = 275\,deg, DE = -27\,deg). These peaks are present in both the full sample and in the reduced sample and coincide with the position of the Gal. centre. Noticeably, the current position of US\,708 does not correspond to the predicted positions of the peaks. The distance distributions (measured from the position of Earth) show a plateau at $\sim 20$\,kpc (this forms a peak when larger distances are included) in the full distribution. This is in contrast to the reduced distribution where the same peak is located at $\sim 190$\,kpc (See App.\,\ref{app:full-D}). These peaks are essentially a reflection of the object number densities being integrated over the entire celestial sphere (compare Fig.\,\ref{fig:radius-initial-conditions}). However, the location of these peaks is instructive of the effect discussed in Sec.\,\ref{sec:galactocentric}, with numbers of lower mass runaways being attenuated due to their higher runaway velocities. If the entire mass range is realized in nature, observations should encounter a relatively higher number of higher mass runaways in the Solar environment than lower mass ones. While there may be a significant observational bias against the detection of higher mass He-stars (see discussion in App.\,\ref{app:observations}), the only observed object being of a lower mass may be an indication that higher mass runaways are selected against. The latter conclusion is significant insofar as donor stars in the He-donor scenario for thermonuclear SNe, particularly in the double detonation scenario, are thought to possess higher masses under most conditions \citep[$M>0.8\Msun$, see][]{NYL2016}. 

\begin{figure*}
	\centering 
	\input{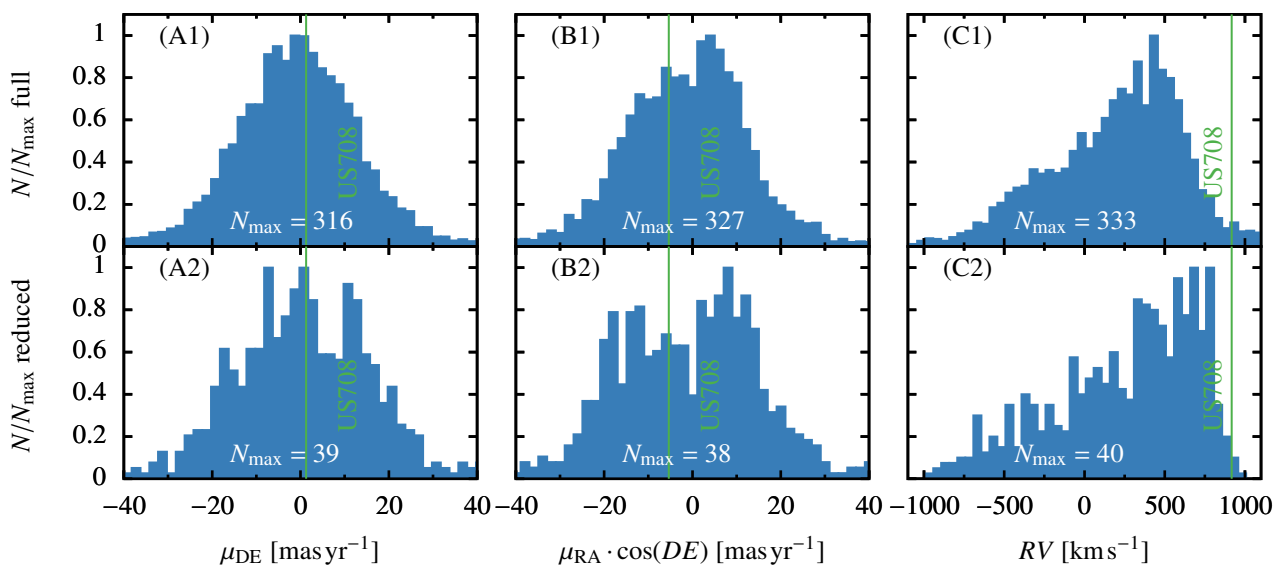}
	\vspace{0.3cm}
	\caption{Parameters of motion of our synthetic population, truncated at a distance of $D=20$\,kpc with proper motions in RA ($\mu_\mathrm{RA}$) and DE ($\mu_\mathrm{DE}$) as labelled. $RV$ is the radial velocity (i.e. velocity along the line of sight).} \label{fig:RV-H}
\end{figure*}

We present the distribution for parameters of motion in Fig.\,\ref{fig:RV-H}. Again we truncate the sample at a distance of $D<20$\,kpc.  We give the parameters of motion for US\,708 for reference. While there are pronounced peaks in the proper motions for the full sample, these peaks are much less pronounced in the reduced sample. In both cases these peaks are located around the zero axis, reflecting objects at greater distances dominating the population. In both cases values of more than $20$\,mas\,yr$^{-1}$ can be expected, mostly representing objects being ejected from the Gal. centre perpendicular to the line of sight. In both cases, the distribution being shifted towards positive velocities is a geometric effect.
Noticeably, there is a pronounced peak in the radial velocities present in both the full sample ($\sim500$\,km$^{-1}$) and the reduced sample ($\sim750-850$\,km$^{-1}$). US\,708 lies in the high-velocity tail of both samples. In any case, we conclude that proper motion is likely not a promising observational filter for these objects with searches for high radial velocities being more attractive.

\section{Conclusions} \label{sec:conclusions}
Combining pre-existing data obtained via detailed stellar evolution calculations, model Gal. potential and density distribution and numerical kinematic calculations, we have constructed a synthetic population of US\,708 analogue HVS from a sample of $10^{6}$ individual ejection trajectories. From this synthetic distribution, we derive predictions for the most likely observational characteristics of members of this hypothetical population. 
Most importantly, we find that, if the entire mass range of ejected He-star masses is encountered in nature, which cannot be reliably inferred from the currently sui generis US 708, there should be a significant excess of higher mass runaways over lower mass ones close to Earth. This prediction is at odds with the only observed object, US\,708, which is likely a lower mass star, This, in itself, contains a discrepancy at least with theoretical models on the double detonation scenario predicting higher mass donor stars. This discrepancy may imply that higher mass donor stars are selected against in the pre-SN evolution of the progenitor binary, in which case all observed runaway He donor stars can be expected to possess masses comparable to that of US\,708.
We further find that high velocity runaways in our sample will primarily be distinguished by high radial velocity, with the probability distribution of the entire sample peaking at $\sim500$\kms, the reduced sample peaking at ($\sim750-850$\kms), with US\,708 located in the high-velocity tail of either distribution. 
We note that US\,708 has been discovered based on a spectrum taken during the Sloan Digital Sky Survey \citep{HHOB2005} and most subsequent searches were based on new releases of SDSS. Hot subdwarfs are most easily detected as faint blue stars at high Galactic latitudes. According to our predictions, the currently available survey area at high Galactic latitudes in the Northern hemisphere turns out to be worst choice to search for sdO/B runaway, because it leaves out both the disk and the bulge. Southern surveys such as SDSS-V and 4MOST will likely be much more suited in this respect.
We conclude that the apparent lack of US\,708 analogues might to a large extend be due to strong observational selection effects. These can be overcome by conducting deep spectroscopic surveys in the South, obtain high-quality spectra of carefully selected candidates and improve the models to fit them, and also by checking the classifications and parameters of potentially misclassified bright candidates from the literature. 

However should future observation campaigns fail to produce further US\,708 analogues the overall viability and prevalence of the SD He donor channel should be seen as seriously challenged on the theoretical side.

\begin{acknowledgements}
PN and ASH would like to thank Holly Preece for useful discussions. ASH thanks the Max Planck Society for support through a Max Planck Research Group. MK is partly supported by Grant No 11521303 and 11733008 of the Natural Science Foundation of China.
\end{acknowledgements}

\bibliographystyle{aa}
\bibliography{paper}

\begin{thebibliography}{55}
\expandafter\ifx\csname natexlab\endcsname\relax\def\natexlab#1{#1}\fi

\bibitem[{{Allen} \& {Santillan}(1991)}]{AS1991}
{Allen}, C. \& {Santillan}, A. 1991, \rmxaa, 22, 255

\bibitem[{{Bauer} {et~al.}(2019){Bauer}, {White}, \& {Bildsten}}]{BWB2019}
{Bauer}, E.~B., {White}, C.~J., \& {Bildsten}, L. 2019, arXiv e-prints,
  arXiv:1906.08941

\bibitem[{{Binney} \& {Tremaine}(2008)}]{2008gady.book.....B}
{Binney}, J. \& {Tremaine}, S. 2008, {Galactic Dynamics: Second Edition}
  (Princeton University Press)

\bibitem[{{Brooks} {et~al.}(2017){Brooks}, {Schwab}, {Bildsten}, {Quataert}, \&
  {Paxton}}]{BSB2017}
{Brooks}, J., {Schwab}, J., {Bildsten}, L., {Quataert}, E., \& {Paxton}, B.
  2017, \apj, 843, 151

\bibitem[{{Brown}(2015)}]{B2015}
{Brown}, W.~R. 2015, \araa, 53, 15

\bibitem[{{Brown} {et~al.}(2009){Brown}, {Geller}, {Kenyon}, \&
  {Bromley}}]{BGK2009}
{Brown}, W.~R., {Geller}, M.~J., {Kenyon}, S.~J., \& {Bromley}, B.~C. 2009,
  \apjl, 690, L69

\bibitem[{{Cappellaro} {et~al.}(1997){Cappellaro}, {Turatto}, {Tsvetkov},
  {Bartunov}, {Pollas}, {Evans}, \& {Hamuy}}]{CTT1997}
{Cappellaro}, E., {Turatto}, M., {Tsvetkov}, D.~Y., {et~al.} 1997, \aap, 322,
  431

\bibitem[{{Di Stefano} {et~al.}(2011){Di Stefano}, {Voss}, \&
  {Claeys}}]{DVC2011}
{Di Stefano}, R., {Voss}, R., \& {Claeys}, J.~S.~W. 2011, The Astrophysical
  Journal, 738, L1

\bibitem[{{Dorsch} {et~al.}(2020){Dorsch}, {Latour}, {Heber}, {Irrgang},
  {Charpinet}, \& {Jeffery}}]{DLH2020}
{Dorsch}, M., {Latour}, M., {Heber}, U., {et~al.} 2020, \aap, 643, A22

\bibitem[{{Evans} {et~al.}(2020){Evans}, {Renzo}, \& {Rossi}}]{ERR2020}
{Evans}, F.~A., {Renzo}, M., \& {Rossi}, E.~M. 2020, \mnras, 497, 5344

\bibitem[{{Figer} {et~al.}(2004){Figer}, {Rich}, {Kim}, {Morris}, \&
  {Serabyn}}]{FRK2004}
{Figer}, D.~F., {Rich}, R.~M., {Kim}, S.~S., {Morris}, M., \& {Serabyn}, E.
  2004, \apj, 601, 319

\bibitem[{{Foley} {et~al.}(2013){Foley}, {Challis}, {Chornock},
  {Ganeshalingam}, {Li}, {Marion}, {Morrell}, {Pignata}, {Stritzinger},
  {Silverman}, {Wang}, {Anderson}, {Filippenko}, {Freedman}, {Hamuy}, {Jha},
  {Kirshner}, {McCully}, {Persson}, {Phillips}, {Reichart}, \&
  {Soderberg}}]{FCC2013}
{Foley}, R.~J., {Challis}, P.~J., {Chornock}, R., {et~al.} 2013, \apj, 767, 57

\bibitem[{{Geier}(2020)}]{G2020}
{Geier}, S. 2020, \aap, 635, A193

\bibitem[{{Geier} {et~al.}(2015){Geier}, {F{\"u}rst}, {Ziegerer}, {Kupfer},
  {Heber}, {Irrgang}, {Wang}, {Liu}, {Han}, {Sesar}, {Levitan}, {Kotak},
  {Magnier}, {Smith}, {Burgett}, {Chambers}, {Flewelling}, {Kaiser},
  {Wainscoat}, \& {Waters}}]{GFZ2015}
{Geier}, S., {F{\"u}rst}, F., {Ziegerer}, E., {et~al.} 2015, Science, 347, 1126

\bibitem[{{Geier} {et~al.}(2019){Geier}, {Raddi}, {Gentile Fusillo}, \&
  {Marsh}}]{GRG2019}
{Geier}, S., {Raddi}, R., {Gentile Fusillo}, N.~P., \& {Marsh}, T.~R. 2019,
  \aap, 621, A38

\bibitem[{{Gnedin} {et~al.}(2002){Gnedin}, {Zhao}, {Pringle}, {Fall}, {Livio},
  \& {Meylan}}]{2002ApJ...568L..23G}
{Gnedin}, O.~Y., {Zhao}, H., {Pringle}, J.~E., {et~al.} 2002, \apjl, 568, L23

\bibitem[{{Heber}(2016)}]{H2016}
{Heber}, U. 2016, \pasp, 128, 082001

\bibitem[{{Heger} \& {Woosley}(2002)}]{HW2002}
{Heger}, A. \& {Woosley}, S.~E. 2002, \apj, 567, 532

\bibitem[{{Hillebrandt} \& {Niemeyer}(2000)}]{HN2000}
{Hillebrandt}, W. \& {Niemeyer}, J.~C. 2000, \araa, 38, 191

\bibitem[{{Hills}(1988)}]{H1988}
{Hills}, J.~G. 1988, \nat, 331, 687

\bibitem[{{Hirsch} {et~al.}(2005){Hirsch}, {Heber}, {O'Toole}, \&
  {Bresolin}}]{HHOB2005}
{Hirsch}, H.~A., {Heber}, U., {O'Toole}, S.~J., \& {Bresolin}, F. 2005, \aap,
  444, L61

\bibitem[{{Irrgang} {et~al.}(2019){Irrgang}, {Geier}, {Heber}, {Kupfer}, \&
  {F{\"u}rst}}]{IGH2019}
{Irrgang}, A., {Geier}, S., {Heber}, U., {Kupfer}, T., \& {F{\"u}rst}, F. 2019,
  \aap, 628, L5

\bibitem[{{Irrgang} {et~al.}(2013){Irrgang}, {Wilcox}, {Tucker}, \&
  {Schiefelbein}}]{IWT2013}
{Irrgang}, A., {Wilcox}, B., {Tucker}, E., \& {Schiefelbein}, L. 2013, \aap,
  549, A137

\bibitem[{{Justham} {et~al.}(2009){Justham}, {Wolf}, {Podsiadlowski}, \&
  {Han}}]{JWP2009}
{Justham}, S., {Wolf}, C., {Podsiadlowski}, P., \& {Han}, Z. 2009, \aap, 493,
  1081

\bibitem[{{Koposov} {et~al.}(2020){Koposov}, {Boubert}, {Li}, {Erkal}, {Da
  Costa}, {Zucker}, {Ji}, {Kuehn}, {Lewis}, {Mackey}, {Simpson}, {Shipp},
  {Wan}, {Belokurov}, {Bland-Hawthorn}, {Martell}, {Nordlander}, {Pace}, {De
  Silva}, {Wang}, \& {S5 collaboration}}]{KBD2020}
{Koposov}, S.~E., {Boubert}, D., {Li}, T.~S., {et~al.} 2020, \mnras, 491, 2465

\bibitem[{{Li} {et~al.}(2003){Li}, {Filippenko}, {Chornock}, {Berger},
  {Berlind}, {Calkins}, {Challis}, {Fassnacht}, {Jha}, {Kirshner}, {Matheson},
  {Sargent}, {Simcoe}, {Smith}, \& {Squires}}]{LFC2003}
{Li}, W., {Filippenko}, A.~V., {Chornock}, R., {et~al.} 2003, \pasp, 115, 453

\bibitem[{{Li} {et~al.}(2020){Li}, {Luo}, {Lu}, {Zhang}, {Li}, {Wang}, {Zuo},
  {Xiang}, {Ting}, {Marchetti}, {Li}, {Wang}, {Zhang}, {Hattori}, {Zhao},
  {Zhang}, \& {Zhao}}]{LLL2020}
{Li}, Y.-B., {Luo}, A.-L., {Lu}, Y.-J., {et~al.} 2020, arXiv e-prints,
  arXiv:2011.10206

\bibitem[{{Liu} {et~al.}(2015){Liu}, {Moriya}, {Stancliffe}, \&
  {Wang}}]{LMSW2015}
{Liu}, Z.-W., {Moriya}, T.~J., {Stancliffe}, R.~J., \& {Wang}, B. 2015, \aap,
  574, A12

\bibitem[{{Liu} \& {Zeng}(2020)}]{LZ2020}
{Liu}, Z.-W. \& {Zeng}, Y. 2020, arXiv e-prints, arXiv:2011.07691

\bibitem[{{Lu} {et~al.}(2010){Lu}, {Zhang}, \& {Yu}}]{LZY2010}
{Lu}, Y., {Zhang}, F., \& {Yu}, Q. 2010, \apj, 709, 1356

\bibitem[{{Luo} {et~al.}(2020){Luo}, {N{\'e}meth}, \& {Li}}]{LNPL2020}
{Luo}, Y., {N{\'e}meth}, P., \& {Li}, Q. 2020, \apj, 898, 64

\bibitem[{{Maoz} {et~al.}(2011){Maoz}, {Mannucci}, {Li}, {Filippenko}, {Della
  Valle}, \& {Panagia}}]{MMF2011}
{Maoz}, D., {Mannucci}, F., {Li}, W., {et~al.} 2011, \mnras, 412, 1508

\bibitem[{{Moehler} {et~al.}(1997){Moehler}, {Heber}, \& {Durell}}]{MHD1997}
{Moehler}, S., {Heber}, U., \& {Durell}, P.~R. 1997, \aap, 317, 83

\bibitem[{{N{\'e}meth} {et~al.}(2016){N{\'e}meth}, {Ziegerer}, {Irrgang},
  {Geier}, {F{\"u}rst}, {Kupfer}, \& {Heber}}]{NZI2016}
{N{\'e}meth}, P., {Ziegerer}, E., {Irrgang}, A., {et~al.} 2016, \apjl, 821, L13

\bibitem[{{Neunteufel}(2020)}]{N2020}
{Neunteufel}, P. 2020, \aap, 641, A52

\bibitem[{{Neunteufel} {et~al.}(2016){Neunteufel}, {Yoon}, \&
  {Langer}}]{NYL2016}
{Neunteufel}, P., {Yoon}, S.-C., \& {Langer}, N. 2016, \aap, 589, A43

\bibitem[{{Neunteufel} {et~al.}(2017){Neunteufel}, {Yoon}, \&
  {Langer}}]{NYL2017}
{Neunteufel}, P., {Yoon}, S.-C., \& {Langer}, N. 2017, \aap, 602, A55

\bibitem[{{Nomoto}(1982)}]{N1982a}
{Nomoto}, K. 1982, \apj, 253, 798

\bibitem[{{Perlmutter} {et~al.}(1999){Perlmutter}, {Aldering}, {Goldhaber},
  {Knop}, {Nugent}, {Castro}, {Deustua}, {Fabbro}, {Goobar}, {Groom}, {Hook},
  {Kim}, {Kim}, {Lee}, {Nunes}, {Pain}, {Pennypacker}, {Quimby}, {Lidman},
  {Ellis}, {Irwin}, {McMahon}, {Ruiz-Lapuente}, {Walton}, {Schaefer}, {Boyle},
  {Filippenko}, {Matheson}, {Fruchter}, {Panagia}, {Newberg}, {Couch}, \&
  {Project}}]{PAG1999}
{Perlmutter}, S., {Aldering}, G., {Goldhaber}, G., {et~al.} 1999, \apj, 517,
  565

\bibitem[{{Press} {et~al.}(1992){Press}, {Teukolsky}, {Vetterling}, \&
  {Flannery}}]{RECIPES_C}
{Press}, W.~H., {Teukolsky}, S.~A., {Vetterling}, W.~T., \& {Flannery}, B.~P.
  1992, {Numerical recipes in C. The art of scientific computing}

\bibitem[{{Raddi} {et~al.}(2020){Raddi}, {Irrgang}, {Heber}, {Schneider}, \&
  {Kreuzer}}]{RIH2020}
{Raddi}, R., {Irrgang}, A., {Heber}, U., {Schneider}, D., \& {Kreuzer}, S.
  2020, arXiv e-prints, arXiv:2011.08862

\bibitem[{{Rossi} {et~al.}(2014){Rossi}, {Kobayashi}, \& {Sari}}]{RKS2014}
{Rossi}, E.~M., {Kobayashi}, S., \& {Sari}, R. 2014, \apj, 795, 125

\bibitem[{{Ruiter}(2020)}]{R2020}
{Ruiter}, A.~J. 2020, IAU Symposium, 357, 1

\bibitem[{{Ruiter} {et~al.}(2009){Ruiter}, {Belczynski}, \& {Fryer}}]{RBF2009}
{Ruiter}, A.~J., {Belczynski}, K., \& {Fryer}, C. 2009, \apj, 699, 2026

\bibitem[{{Schindewolf} {et~al.}(2018){Schindewolf}, {N{\'e}meth}, {Heber},
  {Battich}, {Miller Bertolami}, {Irrgang}, \& {Latour}}]{SNH2018}
{Schindewolf}, M., {N{\'e}meth}, P., {Heber}, U., {et~al.} 2018, \aap, 620, A36

\bibitem[{{Shen} {et~al.}(2018){Shen}, {Boubert}, {G{\"a}nsicke}, {Jha},
  {Andrews}, {Chomiuk}, {Foley}, {Fraser}, {Gromadzki}, {Guillochon}, {Kotze},
  {Maguire}, {Siebert}, {Smith}, {Strader}, {Badenes}, {Kerzendorf}, {Koester},
  {Kromer}, {Miles}, {Pakmor}, {Schwab}, {Toloza}, {Toonen}, {Townsley}, \&
  {Williams}}]{SBG2018}
{Shen}, K.~J., {Boubert}, D., {G{\"a}nsicke}, B.~T., {et~al.} 2018, \apj, 865,
  15

\bibitem[{{Sukhbold} {et~al.}(2016){Sukhbold}, {Ertl}, {Woosley}, {Brown}, \&
  {Janka}}]{SEWB2016}
{Sukhbold}, T., {Ertl}, T., {Woosley}, S.~E., {Brown}, J.~M., \& {Janka}, H.-T.
  2016, \apj, 821, 38

\bibitem[{{Tillich} {et~al.}(2011){Tillich}, {Heber}, {Geier}, {Hirsch},
  {Maxted}, {G{\"a}nsicke}, {Marsh}, {Napiwotzki}, {{\O}stensen}, \&
  {Scholz}}]{THG2011}
{Tillich}, A., {Heber}, U., {Geier}, S., {et~al.} 2011, \aap, 527, A137

\bibitem[{{Vennes} {et~al.}(2017){Vennes}, {Nemeth}, {Kawka}, {Thorstensen},
  {Khalack}, {Ferrario}, \& {Alper}}]{VNK2017}
{Vennes}, S., {Nemeth}, P., {Kawka}, A., {et~al.} 2017, Science, 357, 680

\bibitem[{Waldman {et~al.}(2011)Waldman, Sauer, Livne, Perets, Glasner,
  Mazzali, Truran, \& Gal-Yam}]{WSL2011}
Waldman, R., Sauer, D., Livne, E., {et~al.} 2011, The Astrophysical Journal,
  738, 21

\bibitem[{{Wang} {et~al.}(2013){Wang}, {Justham}, \& {Han}}]{WJH2013}
{Wang}, B., {Justham}, S., \& {Han}, Z. 2013, A\&A, 559, A94

\bibitem[{{Woosley} \& {Kasen}(2011)}]{WK2011}
{Woosley}, S.~E. \& {Kasen}, D. 2011, \apj, 734, 38

\bibitem[{{Xiang} {et~al.}(2018){Xiang}, {Shi}, {Liu}, {Yuan}, {Chen}, {Huang},
  {Wang}, {Wu}, {Tian}, {Huo}, {Zhang}, \& {Zhang}}]{XSL2018}
{Xiang}, M., {Shi}, J., {Liu}, X., {et~al.} 2018, \apjs, 237, 33

\bibitem[{{Yu} \& {Madau}(2007)}]{YM2007}
{Yu}, Q. \& {Madau}, P. 2007, \mnras, 379, 1293

\bibitem[{{Ziegerer} {et~al.}(2017){Ziegerer}, {Heber}, {Geier}, {Irrgang},
  {Kupfer}, {F{\"u}rst}, \& {Schaffenroth}}]{ZHG2017}
{Ziegerer}, E., {Heber}, U., {Geier}, S., {et~al.} 2017, \aap, 601, A58

\end{thebibliography}

\appendix

\section{Remarks on the methodology}

\subsection{Numerical framework} \label{app:SHyRT}
In this study, we use a lightweight stellar kinematics framework ("SHyRT") to compute the ejection trajectories of our sample.
The "SHyRT"-framework (Simulated Hypervelocity Runaway Trajectories) is a newly developed, lightweight numerical tool designed to rapidly compute movement of ejected stars in the Gal. gravitational potential and provide synthetic astrometric observables of the calculated trajectories. At its core, the SHyRT framework solves the basic Newtonian equations of motions in the form
\begin{equation}
\frac{d}{dt} \frac{d\vec{x}}{dt} = - \nabla \Phi(\vec{x}), 
\end{equation}
utilizing a fourth-order Runge-Kutta integration scheme with adaptive step size control as described by \cite{RECIPES_C}. The reliability of this tool was verified by comparison with extant tools designed with the same goal in mind, such as gal\_py and is identical to that utilized by N20.
Importantly, for this study, the SHyRT-framework calculates Gal. rotational velocity curves and includes them in the initial conditions for each simulated trajectory. 

\subsection{Initial conditions} \label{app:initial conditions}
The initial conditions for this simulation consist of seven primary parameters:
Position in three dimensions, the ejection direction in three dimension and the total velocity with respect to the centre of mass of the progenitor system. All of these parameters are randomized as far as possible, using the prescription described below.
For the velocity, we adopt the ejection velocity spectra presented by N20, (pseudo-)randomly\footnote{unless stated otherwise, random numbers were generated using the \texttt{rd()}-function, contained in the \texttt{<random>}-library for c++. } choosing a discrete mass in the range $0.2\Msun \leq M \leq 0.8\Msun$ ($0.05\Msun$ intervals) for the donor/runaway, then randomly choosing a velocity appropriate for that mass according to the spread of the spectrum (see also Appendix\,\ref{app:velocity spectra}). We make a conscious choice to only consider discrete values for runaway masses for convenience in population analysis. The chosen mass range corresponds to the ejection velocity spectra available from N20, the lower limit in that study being dictated by considerations of numerical stability and the upper limit by a desire to explore the production of HVS, with $0.8\Msun$ being close to the upper mass limit for production of HVS in the solar neighbourhood
For the ejection location, we assume that all stars originate from the Gal. plane, so no variation was assumed along the Gal. z-axis (though movement is fully allowed in 3D). For the remaining two dimensions, a polar coordinate system was adopted, and an angle in the range $0-2\,\pi$ was chosen randomly. The radius was also chosen randomly, though weighted such that the line number density of ejection locations obeys $\rho_N(R) \propto \rho(R) \cdot R$, where $\rho_N(R)$ is the number density at distance from the Gal. centre $R$ and $\rho(R)$ \footnote{i.e. the total number of objects is $N = \int_{0}^{\infty} \rho_N(r) \mathrm{dr}$}is the number density at the same location (See App.\,\ref{app:Galactic-density}).
The ejection direction is randomly chosen to conform to a flat isotropic distribution.

\subsection{Ejection velocity spectra} \label{app:velocity spectra}

\begin{figure} 
	\input{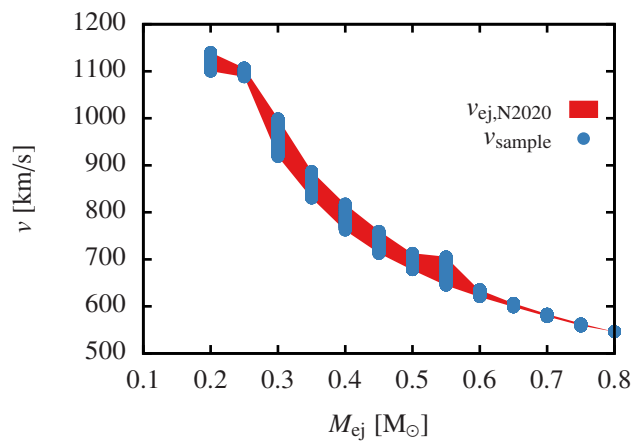}
	\caption{Ejection velocity spectra and sample runaway velocities as used in this study. The red shaded area ($v_\mathrm{ej,N2020}$) denotes the range of ejection velocities as calculated by N20, the blue dots ($v_\mathrm{sample}$) show the derived ejection velocities per system utilized in this study. See also Tab.\,\ref{tab:vvalues}.} \label{fig:vel-spectrum}
\end{figure}

Ejection velocities were chosen according to the ejection velocity spectra calculated by N20 (Figure 8E) in the following way: A mass, assumed to be the runaway's terminal mass before ejection, in the range $0.2\Msun \leq  M_\mathrm{ej} \leq  0.8\Msun$ was chosen randomly. Then, a value in the range $v_\mathrm{ej}(M_\mathrm{ej,min}) - v_\mathrm{ej}(M_\mathrm{ej,max})$ was chosen, again randomly. The result of this process is illustrated in Fig.\,\ref{fig:vel-spectrum}, with exact values for the maximum and minimum ejection velocities for each discrete mass value given in Tab.\,\ref{tab:vvalues}.
The resulting velocity is assumed to be the space velocity of the runaway with respect to the centre of mass of the progenitor binary. 
Once the ejection location was determined (see Sec.\,\ref{sec:assumptions} and App.\,\ref{app:Galactic-density}), the galactocentric ejection velocity is then calculated by taking the local Gal. rotational velocity as well as the, randomly determined, ejection direction into account.

\begin{table} 
	\begin{center}
	\caption{Minimum $v_\mathrm{min}$ and maximum $v_\mathrm{max}$ ejection velocities for runaways of discrete terminal mass $M_\mathrm{ej}$ as provided by N20 and utilized in this study (see also Fig.\,\ref{fig:vel-spectrum}).} \label{tab:vvalues}
	\begin{tabular}{ccc} 
		\hline 
		\hline
		$M_\mathrm{ej}~[\Msun]$ & $v_\mathrm{min}$~[\kms] & $v_\mathrm{max}$~[\kms]  \\  
		\hline
		0.20 & 1100 & 1140 \\
		0.25 & 1088 & 1107 \\
		0.30 & \phantom{0}918   & \phantom{0}999   \\
		0.35 & \phantom{0}830   & \phantom{0}887   \\
		0.40 & \phantom{0}762   & \phantom{0}817   \\
		0.45 & \phantom{0}712   & \phantom{0}760   \\
		0.50 & \phantom{0}678   & \phantom{0}713   \\
		0.55 & \phantom{0}645   & \phantom{0}706   \\
		0.60 & \phantom{0}621   & \phantom{0}636   \\
		0.65 & \phantom{0}599   & \phantom{0}607   \\
		0.70 & \phantom{0}578   & \phantom{0}583   \\
		0.75 & \phantom{0}558   & \phantom{0}563   \\
		0.80 & \phantom{0}547   & \phantom{0}547   \\
		\hline 
	\end{tabular} 
	\end{center}
\end{table}

\subsection{The Galactic density distribution} \label{app:Galactic-density}
As described in Sec.\,\ref{sec:assumptions}, the ejection locations for our model population were randomly chosen, but weighted according to the Gal. density distribution.
The Gal. density distribution was assumed to be static and correspond to Model 1 put forward by \cite{IWT2013} as a revision of \cite{AS1991} in the form
\begin{equation} \label{eq:model-d1.1}
\rho_b(R)= \frac{3 b_b^2 M_b}{4 \pi (R^2+b_b^2)^{5/2}} ,
\end{equation}
for the bulge and
\begin{equation} \label{eq:model-d1.2}
\rho_d(r,z) = \frac{b_d^2 M_d}{4 \pi} \frac{a_d r^2 + \left(a_d + 3 \sqrt{r^2 + z^2}\right)\left(a_d+\sqrt{r^2+z^2}\right)}{\left(z^2+b_d^2\right)^{3/2} \left(r^2+\left(a_d+\sqrt{z^2+b_d^2}\right)\right)^{5/2}}.\end{equation}
for the disc.
See Table \ref{tab:modelp} for utilized parameter values.
Note that, while \citet{IWT2013} do provide a density distribution for the Gal. dark matter halo, the initial population in this study is assumed to be unaffected by the dark matter halo, only corresponding to the density distribution of baryonic matter.

The radial distribution of ejection location is generated as follows: 
Starting from the outer edge of the distribution (i.e. the edge of the Gal. disc), the radial coordinate is subdivided into a (not predetermined) number of bins. Starting at a relative bin width ($W_\mathrm{cbin}$) of unity, every subsequent bin is assigned a width such that $W_\mathrm{cbin} \propto 1/(\rho(R) \cdot R)$. Radial coordinates are then assigned by randomly choosing one bin, then randomly choosing a value for $R$ within that bin. We allow radii up to 200\,kpc.

\begin{figure} 
	\input{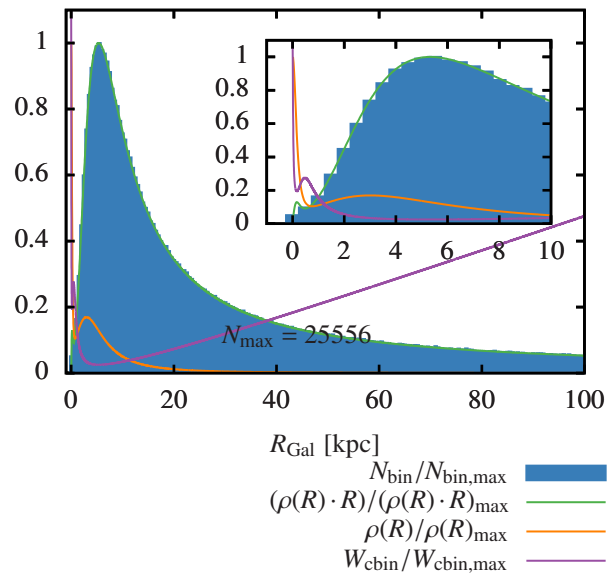}
	\caption{Normalized histogram of the radial distribution of ejection locations, overlaid with the normalized line density $\rho(R) \cdot R$, normalized mass density (see Eqs.\,\ref{eq:model-d1.1} and \ref{eq:model-d1.2}). $W_\mathrm{cbin}/W_\mathrm{cbin,max}$ is the normalized width of the computational bins (see text). The inset is the same as the main plot, but restricted to the inner 10\,kpc.} \label{fig:radius-initial-conditions}
\end{figure}

\begin{figure} 
	\input{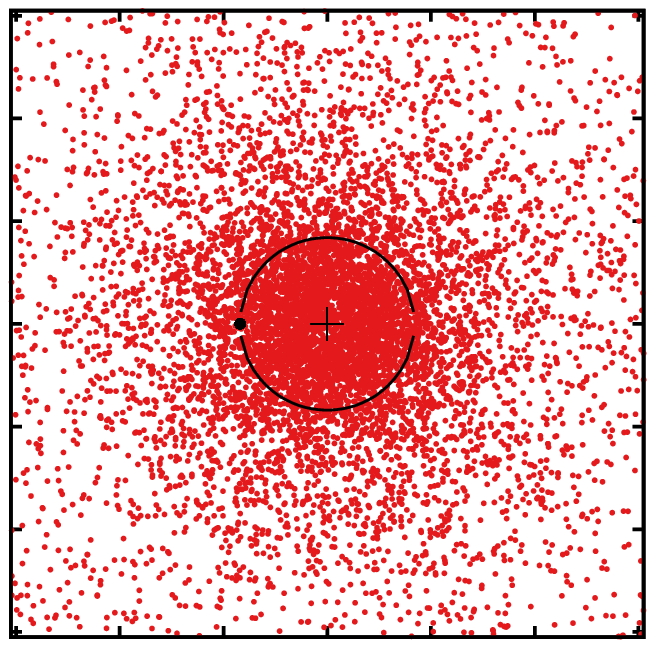}
	\caption{Illustration of randomized ejection locations used in this study in Galactocentric Cartesian coordinates. Note that, for legibility, only every 100th location is plotted. The black circle indicates the orbital radius of the Solar system around the Gal. centre. The positions of Earth (black dot) and the Gal. centre (black cross) are indicated.} \label{fig:ejloc-xy}
\end{figure}

The radial distribution of ejection location is shown in Fig.\,\ref{fig:radius-initial-conditions}. We deem the randomized distribution to adequately conform to expectations. The distribution of ejection locations in the x-y-plane is shown in Fig.\,\ref{fig:ejloc-xy}. We note in passing, that the most likely ejection location of the observed object US\,708 lies close to the maximum in line density at $\sim 6\,\mathrm{kpc}$.

\begin{table} 
	\caption{Parameters used in Eqs.~\ref{eq:model-d1.1}-\ref{eq:model-d1.2} and Eqs.~\ref{eq:model1.1}-\ref{eq:model1.4}.} \label{tab:modelp}
	\begin{tabular}{ccccc} 
		\hline 
		\hline
		& $M_\mathrm{b/d/h}~[\mathrm{M}_\mathrm{G}]$ & $a_\mathrm{d/h}~[\mathrm{kpc}]$ & $b_\mathrm{b/d}~[\mathrm{kpc}]$ & $\Lambda~[\mathrm{kpc}]$  \\  
		\hline
		Bulge$_\mathrm{b}$& $409\pm63$  &  & $0.23\pm0.03$  &\\ 
		Disk$_\mathrm{d}$ & $2856^{+376}_{-202}$  & $4.22^{+0.53}_{-0.99}$ & $0.292^{+0.020}_{-0.025}$ &\\ 
		Halo$_\mathrm{h}$ & $1018^{+27933}_{-603}$ & $2.562^{+25.963}_{-1.419}$ &   &$200^{+0}_{-82}$\\ 
		\hline 
	\end{tabular} 
\end{table}

\subsection{The Galactic potential} \label{app:Galactic-potential}

The functional form of the Gal. potential corresponds to the density distribution as described above \citep[Model I, as proposed by ][]{IWT2013}:
\begin{equation} \label{eq:model1.1}
\Phi_b(R)= -\frac{M_b}{\sqrt{R^2+b_b^2}}
,\end{equation}
for the bulge, where $R$ is the distance from the Gal. centre,
\begin{equation}
\Phi_d(r,z)= -\frac{M_d}{\sqrt{r^2+(a_d+\sqrt{z^2+b_d^2})}}
,\end{equation}
for the disk, where $r$ is the distance from the Gal. Center in the $x$-$y$-plane and $z$ is the distance from the $x$-$y$-plane and
\begin{align} \label{eq:model1.4}
\Phi_h(R)&= -\frac{M_h}{a_h}\left[ \frac{1}{\gamma -1} \ln\left( \frac{1+(R/a_h)^{\gamma-1}}{1+(\Lambda/a_h)^{\gamma-1}} \right) - \frac{(\Lambda/a_h)^{\gamma-1}}{(1 +\Lambda/a_h)^{\gamma-1}} \right]~\mathrm{if}~R < \Lambda \nonumber \\
&= -\frac{M_h}{R} \frac{(\Lambda/a_h)^{\gamma-1}}{(1 +\Lambda/a_h)^{\gamma-1}}~\mathrm{if}~R \geq \Lambda, 
\end{align}
for the dark matter halo, with $\gamma=2$, $G=1$ and the other parameters given in Table \ref{tab:modelp}.
The population is assumed to be non-self-gravitating. Therefore each simulated star is an isolated object moving in a static Gal. potential corresponding to the model presented above. 

\subsection{Runaway lifetimes and luminosities} \label{app:lifetimes}
The state of the former donor star at the point of ejection is, unfortunately, not accessible to our methodology. This leads to the necessity for adopting a number of compromises in the prediction of certain observationally useful parameters, as will be explained, and the remedies adopted, in this appendix. The state of the runaway at ejection is not accessible to us for a number of reasons:
\begin{enumerate}
	\item The ejection velocity spectra adopted from N20 allowed conclusion that the terminal state and structure of the donor (as long as it is core He burning) has less influence on the achievable runaway velocity than its mass. This is not true in for the aftermath of the SN ejection, as the amount of available core He determines the length of the runaways remaining He MS lifetime. It is therefore to be expected that, unless prevented by some unrelated mechanism, two runaways, ejected by the same mechanism, may, despite having identical masses, have a significant difference in remaining He MS lifetimes.
	\item The SN event itself may significantly alter the structural and observational properties of the runaway star. As shown, most recently for this scenario, by \citet{BWB2019}, the donor star is expected to lose a significant amount of material (more than 50\% of its pre-explosion mass) and accumulate enough entropy from the shock interaction to be expand to a structure not unlike that of a protostar. The ejected star may, after this, have lost enough mass to become incapable for further He fusion (generally, hydrogen depleted objects below about $0.3\Msun$ are incapable of He burning). If so, the ejected object will form first a proto-WD, then, after further contraction, a (extremely-) low mass WD. There is, at the time of writing insufficient data available in the literature to cover the entirety of our parameter space and all eventualities.
\end{enumerate}
In order to confront these issues we adopt a the following assumptions and prescriptions:
We neglect the difference in remaining He MS lifetimes of our runaways. Instead, we adopt certain remaining lifetimes according to terminal He donor mass, corresponding to the nuclear timescale of a typical He-star of the same mass. This implies that at, ejection, each runaway has the entirety of its normal He burning lifetime ahead of it (except if $M\leq0.3\Msun$, as discussed below). This leads to an underestimation of each star's luminosity, as most stars of the type under consideration here are expected to brighten over time, but an overestimation of stars in their observable phase since, as mentioned above, most of the objects here will have a diminished remaining He burning lifetime at the point of ejection. The assumed lifetimes are given in Fig.\,\ref{fig:helife}. Here, as stated, we assume a lifetime as given by the star's nuclear lifetime if $M>0.3\Msun$ and thermal collapse on a timescale \citep[corresponding to the values derived by][]{BWB2019} of 10\,Myr if  $M<0.3\Msun$.

\begin{figure} 
	\begin{center}
	\input{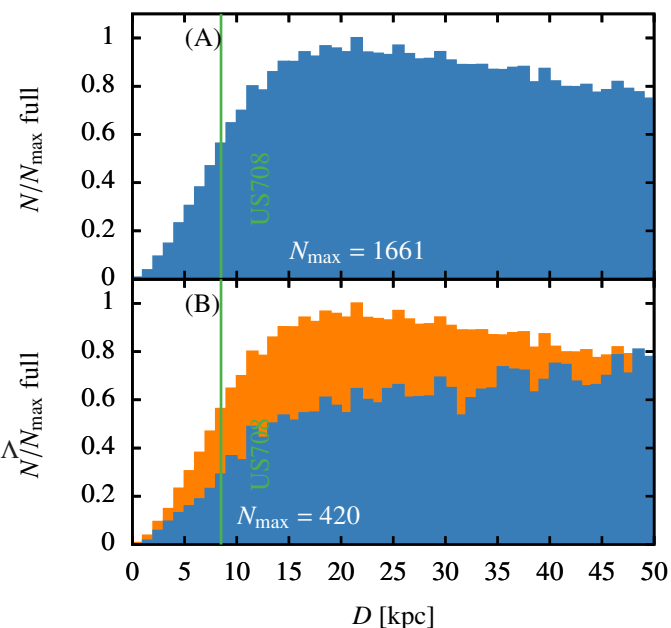}
	\caption{Same as Fig.\,\ref{fig:obs-coord-RADE} C but with artificially reduced lifetimes, i.e.$\tau \rightarrow \tau/2$.} \label{fig:RADE-ref}
	 \end{center}
\end{figure}

\begin{figure} 
	\input{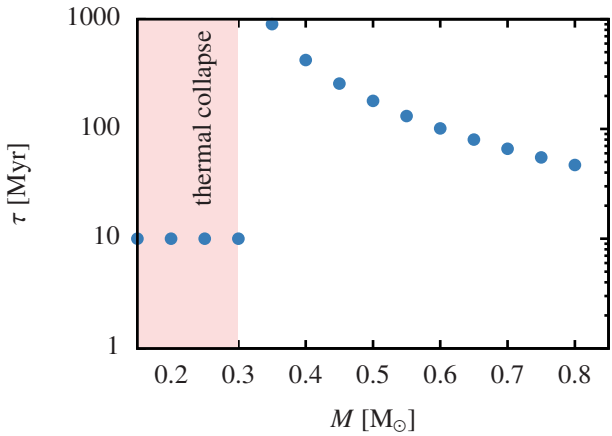}
	\caption{Assumed mass-dependent lifetimes of ejected stars in this study. Stars of $M<0.3\Msun$ are assumed to collapse and form (proto-) WDs without any further He burning.} \label{fig:helife}
\end{figure}

In order to test our assumptions, we redo part of our analysis by artificially reducing the lifetimes of the ejected objects by half. We find that, beyond leading to a more pronounced peak at the same place in the distance distribution, which is a result of an expectedly more pronounced over-density in the vicinity of the initial distribution, this does not significantly alter our results (Fig.\,\ref{fig:RADE-ref}).

%Due to the aforementioned uncertainties concerning to post-ejection evolution of the runaways, a detailed discussion of the Luminosity-effective Temperature evolution of our sample is not deemed worthwhile at the moment. Instead, we adopt a simplified scheme, calculating the absolute magnitude of each runaway based on its mass by extrapolating from the observational properties of US\,708 through homology relations, namely:
%\begin{equation} \label{eq:hom-rel}
%\frac{L_1}{L_2} = \left( \frac{M_1}{M_2} \right)^3 \left( \frac{\mu_1}{\mu_2} \right)^4
%\end{equation}
%with $M$ the mass, $L$ luminosity and $\mu$ the mean atomic weight, which is assumed to remain unchanged in this instance \citep[see e.g.][]{STELLAR_STRUCTURE_AND_EVOLUTION}.

\section{Effects of dynamical friction} \label{app:dyn-friction}

Runaway sdO/B stars can be formed in dense stellar systems such as globular clusters, mediated by dynamical interactions \citep[as reported by][]{MHD1997}. A detailed study of runaway sdO/B stars from these systems is beyond the scope of this paper. However, we do remark that, since the ejection velocity is mostly dictated by the donor's mass prior to detonation, a runaway sdO/B star in, e.g., a globular cluster is expected to have similar ejection velocity compared to the field. Furthermore, given that the escape speed of globular clusters is up to $\sim 100\,\mathrm{km\,s^{-1}}$ \citep{2002ApJ...568L..23G} and the ejection speed lies roughly in the range between 550 and 1150 $\mathrm{km\,s^{-1}}$ (cf. Fig.~\ref{fig:vel-spectrum}), we expect that the internal dynamics of the cluster have negligible impact on the trajectory of the runaway star as it escapes the cluster. 

More quantitatively, encounters with other stars in the cluster as the runaway star moves through it can be described to first approximation as a dynamical friction (DF) process\footnote{Here, we ignore the possibility of interactions with binary or higher-order stars.}. The timescale $T_\mathrm{DF}$ for DF to change the runaway's speed $v$ by order itself can be estimated as (\citealt{2008gady.book.....B}, S8.1) 
\begin{align}
\label{eq:tDF}
\nonumber T_\mathrm{DF} &\equiv \left ( \frac{1}{v} \frac{\mathrm{d} v}{\mathrm{d} t} \right )^{-1} \sim \frac{v^3}{4\pi G^2 M \rho \ln \Lambda} \left [ \mathrm{erf}(X) - \frac{2X}{\sqrt{\pi}} \exp \left (-X^2 \right ) \right ]^{-1} \\
\nonumber &\approx \frac{v^3}{4\pi G^2 M \rho \ln \Lambda} \\
\nonumber & \approx 1.3 \times 10^4 \,\mathrm{Gyr} \, \left ( \frac{v}{600 \,\mathrm{km\,s^{-1}}} \right )^3 \left ( \frac{M}{0.4 \, \Msun} \right )^{-1} \left ( \frac{\rho}{10^4\,\Msun \,\mathrm{pc^{-3}}} \right )^{-1} \\
&\quad \times \left ( \frac{\log \Lambda}{18} \right )^{-1},
\end{align}
where $X \equiv v/(\sqrt{2} \sigma)$ and $\sigma$ is the velocity dispersion, $\rho$ is the stellar density, and $\Lambda$ the Coulomb factor which we estimate as $\Lambda \sim b_\mathrm{max}/b_{90} \sim b_\mathrm{max} v^2/[G(M + m_\star)]$ with $b_\mathrm{max}$ the maximum impact parameter and $m_\star$ the typical stellar mass. Assuming $M = 0.4 \Msun$, $m_\star = 1 \, \Msun$, $v = 600\,\mathrm{km\,s^{-1}}$, and $b_\mathrm{max} = 1\,\mathrm{pc}$, $\log \Lambda \sim 18$. The second line in Eq.~\ref{eq:tDF} applies since $X \sim 42 \gg 1$ assuming $\sigma = 10\,\mathrm{km\,s^{-1}}$ (e.g., \citealt{2002ApJ...568L..23G}). The numerical estimate in Eq.~\ref{eq:tDF} shows that $T_\mathrm{DF}$ greatly exceeds the crossing or even the Hubble time for the smallest relevant ejection velocities, indicating that DF is unimportant. This is also borne out by a number of numerical direct $N$-body integrations that we carried out of $N=10^3$ to $10^4$ stars in a Plummer sphere with a Salpeter mass distribution, assigning an ejection velocity to one star in the centre. The latter's resulting trajectory is virtually indistinguishable from the analytic expectation within the Plummer potential, and any deviation decreases with increasing $N$.  

\section{Observational considerations} \label{app:observations}

The distinction between HVS and fast halo stars is, besides kinematics, as discussed above, made even more complicated by the fact that hot subdwarfs are affected by diffusion and mixing processes in their atmospheres \citep{SNH2018,DLH2020}, which change their primordial abundances and make chemical tagging as population tracer not feasible. The fraction of bound runaways and halo stars high above the Galactic plane is therefore hard to estimate. 

Another selection bias is provided by the magnitude limit of the current surveys. To obtain RVs and in particular atmospheric parameters of sufficient quality for a spectrophotometric distance determination (as provided by G15), current instrumentation only allows to observe stars up to distances of about 10\,kpc. Our simulations show, however, that most runaways are found much farther away from us.

The apparent lack of higher mass He-stars with 0.8\,Msun might also be explained by selection effects. In contrasts to main sequence stars, which have a well-defined empirical mass-radius relation, the masses of hot subdwarfs are not as easy to determine and the theoretically predicted core-mass for the He-flash in low-mass red giants is usually adopted as canonical mass. Most recently, accurate parallaxes from Gaia DR2 became available and can be used along with high-quality atmospheric parameters and photometry to constrain to mass \citep{IGH2019}. For helium-enriched stars such as US\,708, however, the systematic uncertainties on crucial parameters such as the surface gravity are still quite high and make a precise mass determination difficult. Given these problems, the yet missing population of intermediate-mass runaways might just be hidden among the already discovered stars.

And finally, intermediate-mass helium stars can be quite luminous and can therefore be detected far away. Since they are predicted to be close to the disk, the substantial reddening caused by interstellar matter in the disk will then change their apparent colours significantly and move them out of the search spaces for our survey \citep{GRG2019}. Furthermore, such stars can be easily misclassified as main sequence stars as in the famous case of the proposed black hole binary LB-1 \citep{IGH2019}. 

\section{Supplementary results} \label{app:full-D}
We provide an extended heliocentric distance distribution, omitted from the main body of the paper for space, in this appendix.
\begin{figure} 
	\begin{center}
	\input{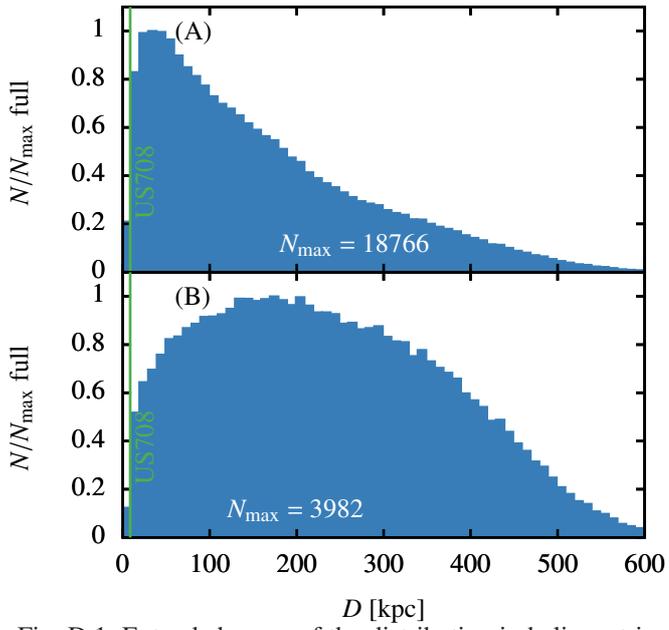}
	\caption{Extended range of the distribution in heliocentric distance ($D$). Upper panels show the entire mass range (full sample) of runaway masses, lower panels show the same for the US\,708-analogue subsample (reduced sample). The heliocentric distance of US\,708 is as indicated.} \label{fig:D-single}
	\end{center}
\end{figure}

\end{document}